\documentclass[3p,times,procedia]{elsarticle}
\usepackage[bookmarks=false]{hyperref}
    \hypersetup{colorlinks,
      linkcolor=blue,
      citecolor=blue,
      urlcolor=blue}

\usepackage{amssymb}

\biboptions{authoryear}

\usepackage[figuresright]{rotating}

\begin{document}
\begin{frontmatter}



\title{Simulation of mechanical effects of hydrogen in bicrystalline Cu using DFT and bond order potentials}


\author{Vasileios Fotopoulos$^{1,*}$}
\author{Alexander L. Shluger$^{1}$}

\address{$^{1}$Department of Physics and Astronomy, University College London, Gower Street, London WC1E 6BT, UK.}

\begin{abstract}
 Hydrogen embrittlement is a prime cause of several degradation effects in metals. Since grain boundaries (GBs) act efficiently as sinks for hydrogen atoms, H is thought to segregate in these regions, affecting the local formation of dislocations. However, it remains unclear at which concentrations H begins to play any role in the mechanical properties of Cu. In the current study, we use density functional theory (DFT) to assess the accuracy of a bond order potential (BOP) in simulating the segregation of H in Cu $\Sigma$5 GB. BOP accurately predicts the most favorable segregation sites of H in Cu GB, along with the induced lattice relaxation effects. H is found to weaken the crystal by reducing the GB separation energy. Classical molecular dynamics (MD) simulations using BOP are performed to evaluate the concentration of H in bicrystalline Cu required to substantially impact the crystal's mechanical strength. For concentrations higher than 10 mass ppm, H significantly reduces the yield strength of bicrystalline Cu samples during uniaxial tensile strain application. This effect was attributed to the fact that H interstitials within the GB promoted the formation of partial dislocations. 
\end{abstract}

\cortext[cor]{Corresponding author.}
\end{frontmatter}

\email{vasileios.fotis.19@ucl.ac.uk}



\section{Introduction}
\label{Introduction}

Hydrogen is known to degrade the mechanical performance of metals~\cite{troiano1960role}, a phenomenon referred to as H embrittlement. Though the impact of H embrittlement on performance of electronic devices and industrial components is known to be detrimental, the exact mechanism of this effect is not fully understood~\cite{louthan2008hydrogen, djukic2016hydrogen, li2019effects}. The origin of various complex defects, such as voids and cracks, which have been shown to initiate at grain boundaries (GBs) of polycrystalline metals~\cite{moser2021electropolishing,konishi2002effect} is often associated with the presence of non-metallic impurities, such as H~\cite{yamaguchi2019first}, which tend to segregate at these regions~\cite{lousada2022hydrogen}. Once introduced in the GBs, H interstitials can have a stabilizing effect on metal vacancies~\cite{fotopoulos2023molecular}, facilitating the initiation of voids. Apart from GBs, dislocations are also known to act as trapping centers for H~\cite{wan2019hydrogen, lynch2012hydrogen}. The absorption of H atoms by either dislocations or grain boundaries can promote corrosion and loss of ductility via weakening of metal-metal bonds~\cite{ birnbaum1994hydrogen, robertson2001effect}.

Electrochemically deposited (ECD) Cu films are commonly used as an interconnect in electronic devices, which operate at elevated temperatures or under high-stress conditions~\cite{articleNeedleman}. H is introduced during the deposition through aqueous electrolytes~\cite{gabe1997role}. Once introduced, atomic H can diffuse through the film, getting trapped and  aggregating in defects~\cite{li2020review}. Compared to other metals, such as Ni and Pd~\cite{wipf1997diffusion}, the amount of incorporated H in Cu is significantly lower~\cite{fotopoulos2023thermodynamic}. Thus the amount of H required to significantly impact the mechanical properties of Cu is unclear. This can be elucidated via theoretical simulations.

Simulating the mechanical properties of polycrystalline metals using atomistic methods requires large scale models, beyond those attainable to \textit{ab initio} approaches~\cite{yoo2021density}. Therefore many previous studies employed empirical potentias to model metallic systems under external strain~\cite{wang2013defect}. Bond order potentials (BOPs) have been shown to accurately predict bulk properties of not only pure metals but also of various Cu~\cite{zhou2015analytical} and Al hydrides~\cite{zhou2018bond}. However, the properties of H interstitials in Cu GBs along with the induced relaxation effects obtained via such potentials have not yet been compared with DFT results. Therefore, how accurately BOP can predict the properties of individual H atoms in Cu GBs remains unclear.

Here we provide a more in-depth investigation into the mechanical effect of H in bicrystalline Cu. Comparison between the density functional theory (DFT) and BOP calculations demonstrates that BOP can accurately predict favorable segregation sites and the lattice relaxation induced by H in Cu $\Sigma$5 GB. Our DFT results show that the presence of H increases the separation energy of Cu GB. The latter is attributed to the induced strong lattice relaxation and significant charge redistribution due to the presence of H. Moreover, H in bicrystalline Cu is found to reduce the yield strength of the crystal at all tested concentrations, however, for a substantial reduction, a minimum of 10 mass ppm of interstitial H atoms in the GB is required. These effects are attributed to the fact that H promotes the emission of partial and Shockley dislocations from the grain boundaries during uniaxial tensile strain application. 

The paper is organized as follows. In section 2, the computational parameters of our static calculations using DFT and BOP are outlined along with the technical details of BOP MD simulations. In Section 3.1, we compare the DFT and BOP static results of H in small Cu GB systems. Our ab initio results showcasing the role of H in Cu GB under strain are presented in Section 3.2. In Section 3.3, the impact of different concentrations of H atoms on the mechanical strength of larger bicrystalline Cu models under tensile strain is presented. Finally, the main conclusions are summarized in Section 4.

\section{Methods}

\subsection{Bond Order Potentials }
 
BOPs differ from empirical force fields, such as the embedded atom method (EAM) potentials, in that the covalent bonding and charge transfer are contained within the expansion of the total energy. Thus, BOPs can be applicable to a wide range of materials and types of bonding. BOPs are derived from DFT via the simplification that the tight-binding (TB) approximation provides~\cite{drautz2015bond, silver1996kernel, voter1996linear, drautz2015bond} and allow us to model the  Cu--Cu, Cu--H and H--H interactions. For the Cu-H interactions, crucial is the accurate description of $\sigma$ bonding. $\pi$ bonding is not included and the potential uses as a basis the potentials developed by Pettifor and Tersoff~\cite{oleinik1999analytic}, which are considered to describe accurately the $\sigma$ bonding between metals and hydrogen atoms~\cite{zhou2015analytical, juslin2005analytical, zhou2018bond}. It is important to emphasize here that these potentials have been developed to accurately describe the growth of Cu in a H-rich environment. Thus, since in our case the goal is to investigate the H segregantion in bicrystalline Cu, comparison with DFT results is needed.

\subsection{\label{sec:level3}Computational Details: First Principles}

\begin{figure}[t]\vspace*{4pt}
 \centering
   \begin{tabular}{cc}
      \hspace{4mm}\textbf{(a)} & \hspace{5mm}\textbf{(b)}\\

   \includegraphics[scale=0.035]{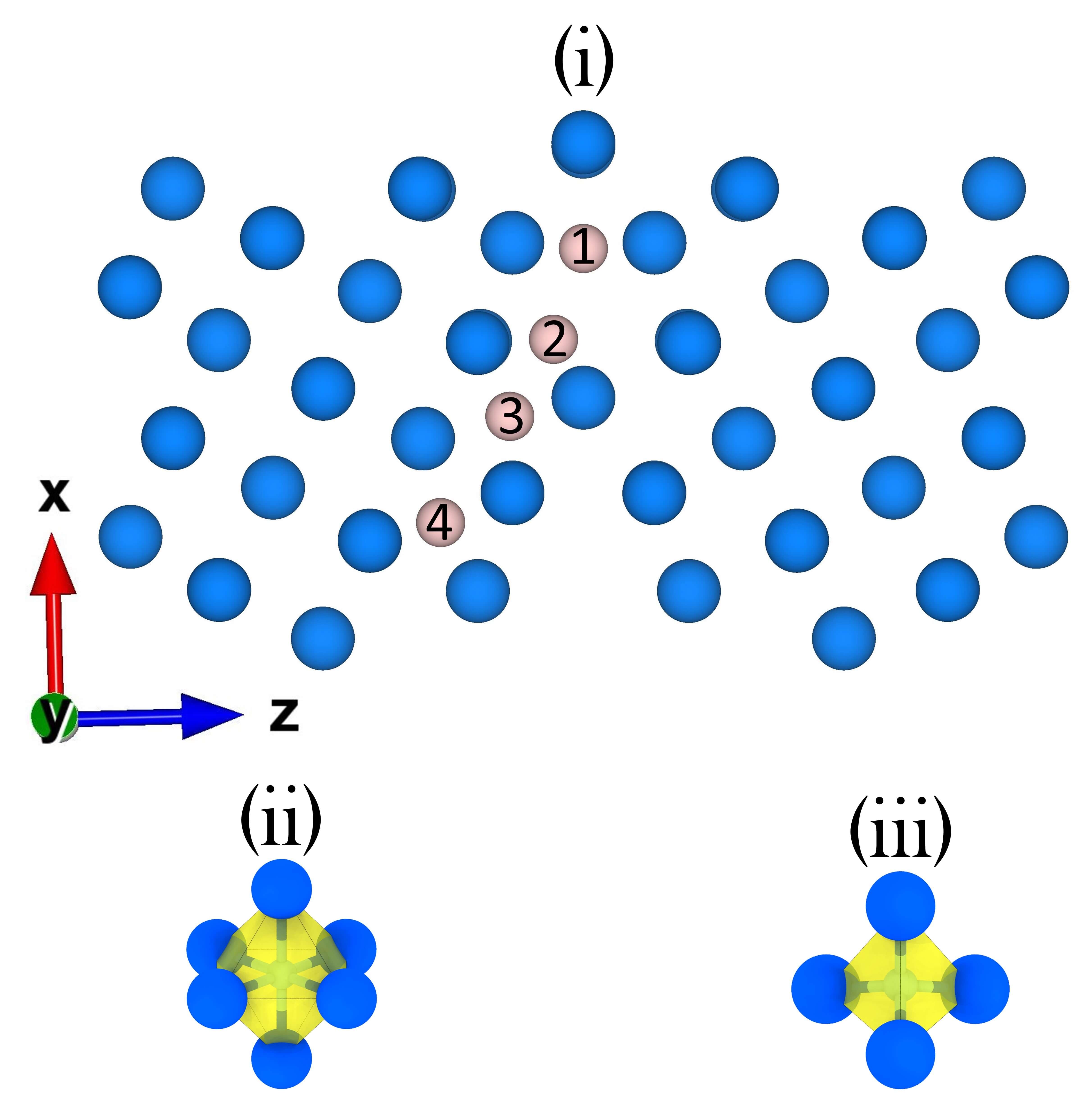}&
      \includegraphics[scale=0.025]{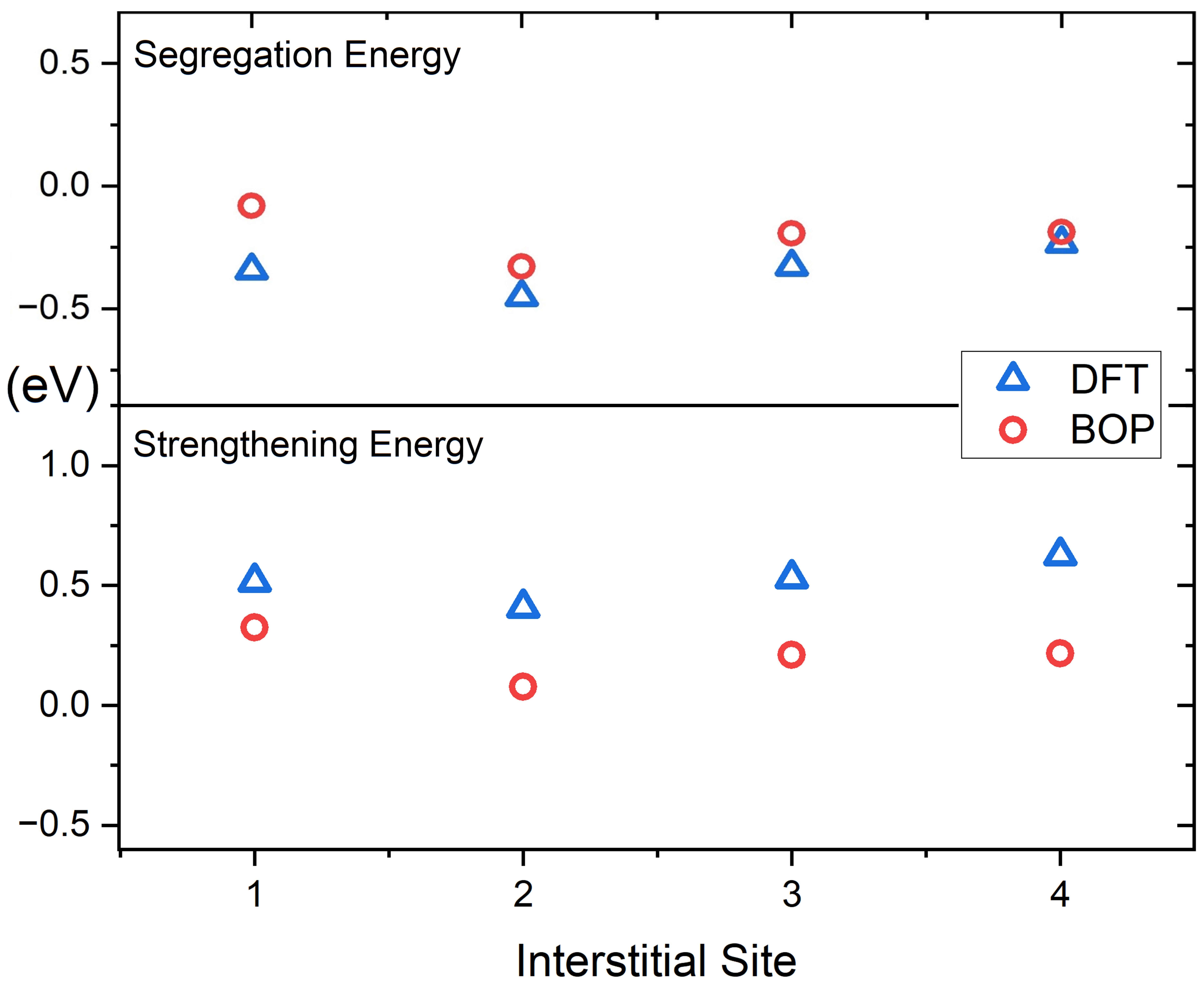}\\
\end{tabular}

\caption{(a)  (i) Shows the examined interstitial H sites (grey spheres) inside the Cu $\Sigma$5 GB; (ii) octahedral and (iii) tetrahedral sites in the bulk. (b) Comparison of segregation and strengthening energies computed using DFT and BOP. To visualize the resulting DFT configurations and charge distributions, VESTA is used~\cite{momma2008vesta}.}
  \label{fig:hingrainboundaries}
\end{figure}

Static DFT calculations are carried out for 76-atom (GB; 7.27\,{\AA}$\times$8.13\,{\AA}$\times$24.39\,{\AA} cell dimensions) and 108-atom (bulk; 10.86\,{\AA}$\times$10.86\,{\AA}$\times$10.86\,{\AA} cell dimensions) periodic simulation cells. The Vienna Ab Initio Simulation Package (VASP)~\cite{kresse1993ab, kresse1996efficient,kresse1996efficiency} and the Perdew–Burke–Ernzerhof (PBE) GGA functional~\cite{perdew1996generalized} are used. In line with previous studies in Cu~\cite{ganchenkova2014effects}, a mixture of the~\cite{er1975iterativecalculationof} and RMM-DIIS~\cite{pulay1980convergence,wood1985new} algorithms are used to minimize the energy with an energy tolerance of 10$^{-5}$\,eV. We consider the (210)[100] $\Sigma5$ twin boundary (see Figure \ref{fig:hingrainboundaries}(a)(i)), which is one of the lowest energy grain boundaries in Cu~\cite{wu2016first}. The simulation cell is periodically translated in the x, y, and z directions. Along the z direction, a 10\,{\AA} vacuum is added in order to avoid interactions between periodically translated images. For the 76-atom GB and the 108-atom bulk cells, in line with previous works~\cite{nazarov2012vacancy, nazarov2014ab, bodlos2022energies}, converged 5$\times$4$\times$1 and 4$\times$4$\times$4 k-point grids are used, respectively, with an energy cutoff of 450\,eV. In the case of the GB cells, due to the added vacuum,  1 k-point is used along the z direction. The Cu pseudopotential with 11 valence electrons ($3d^{10} 4s^1$) is used in all calculations.

\subsection{Computational Details: Bond Order Potentials}

All BOP simulations are conducted using the Large-scale Atomic Molecular Massively Parallel Simulator (LAMMPS) code~\cite{thompson2022lammps}. Geometry optimization is performed using the conjugate gradient (CG) algorithm with a force tolerance of $10^{-10}$\,eV/{\,\AA}. A further equilibration process is followed under an isothermal-isobaric NPT ensemble while retaining a constant temperature of 0\,K for 0.02\,ns. A timestep of 1\,fs is chosen for all MD simulations. For our uniaxial tensile strain simulations, a periodic in all 3 directions 8.1\,nm$\times$24.3\,nm$\times$3.6\,nm (210)[100] $\Sigma5$ 120,000-atom grain boundary cell is used. H interstitial atoms are introduced in the grain boundary region of the bicrystalline simulation cell at concentrations ranging from 10 to 40 mass ppm. The mentioned concentrations are within the reported experimental range~\cite{fukai2003superabundant, fukumuro2011influence}.

During tensile loading, uniaxial tensile deformations are introduced along the y-axis at a constant strain rate of 10$^8$\,s$^{-1}$ and a constant temperature of 300\,K, in line with previous MD studies in Cu~\cite{zhao2016ductile,zhou2017molecular}.  During the simulations, the x and z-direction boundaries are allowed to vary to maintain the corresponding components of the stress equal to zero by using the isobaric-isothermal NPT ensemble. For determining the atomic positions and velocities, the Verlet numerical integration algorithm is used~\cite{kapoor1998determination}. The formed dislocations are identified using the dislocation extraction algorithm (DXA)~\cite{stukowski2012automated}. The common neighbor analysis (CNA) is used for the structural analysis.

\subsection{Energetic Parameters}

Segregation energies indicate whether H prefers to sit at grain boundaries or in the bulk. In order to calculate the segregation energies of H in Cu GB, four interstitial sites are tested (see Figure \ref{fig:hingrainboundaries}(a)(i)). The selection of sites is based on previous DFT segregation studies in Cu GBs~\cite{razumovskiy2018solute,wurmshuber2022mechanical}. Two interstitial sites in the bulk of bulk Cu are used as a reference with Figures \ref{fig:hingrainboundaries}(a)(ii) and (iii) showing the tested octahedral and tetrahedral sites in the bulk, respectively. The segregation energies are computed using the following formula:

\begin{equation}
    E_{seg}=(E_{GB+H}-E_{GB}-(E_{Bulk +H}-E_{Bulk},
    \label{eq:impurities}
    \end{equation} 
    
\noindent{where} E$_\mathrm{Bulk+H}$ and E$_\mathrm{GB+H}$ are the minimized total energies of the 108-atom bulk and 76-atom grain boundary Cu cells, respectively, containing one H interstitial atom. E$_\mathrm{Bulk}$ and E$_\mathrm{GB}$ are the respective H-free bulk and grain boundary energies. Negative energies correspond to favorable GB segregation. 

The impact of impurities on the grain boundary strength can be measured by the strengthening energy (E$_\mathrm{str}$) based on the Rice-Wang~\cite{rice1989embrittlement} model:

\begin{equation}
    E_{str}=(E_{GB+H}-E_{GB}-(E_{Surf+H}-E_{Surf},
    \label{eq:impurities}
    \end{equation} 

\noindent{where} E$_\mathrm{Surf}$ and E$_\mathrm{Surf+H}$ are the total energies of the H-free Cu free surface and free surface with one H atom. For surface simulations, a 3$\times$3$\times$3 108-atom (100) simulation cell is used with an extended z dimension (added vacuum), having a thickness of 10.86\,{\AA}. H atoms are introduced at the hollow sites of the surface. A negative value of strengthening energy means that the impurity will enhance the grain boundary strength, while a positive value suggests weakening effect.

For the DFT tensile strain simulations, we employ the approach proposed by~\cite{yamaguchi2005grain}. Based on the latter, a fracture plane is chosen in advance, and the pre-crack is introduced along this plane as the separation between the two grains is increased. Then, the separation energies are simply computed as the difference between the total energy of the GB cell at a specific separation between the two grains and the total energy of the GB cell at the equilibrium separation. The choice of the fracture plane is based on previous DFT studies of embrittlement effects in Fe~\cite{scheiber2020influence}.

\section{Results}

\begin{figure}[t]\vspace*{4pt}
 \centering
\includegraphics[scale=0.03]{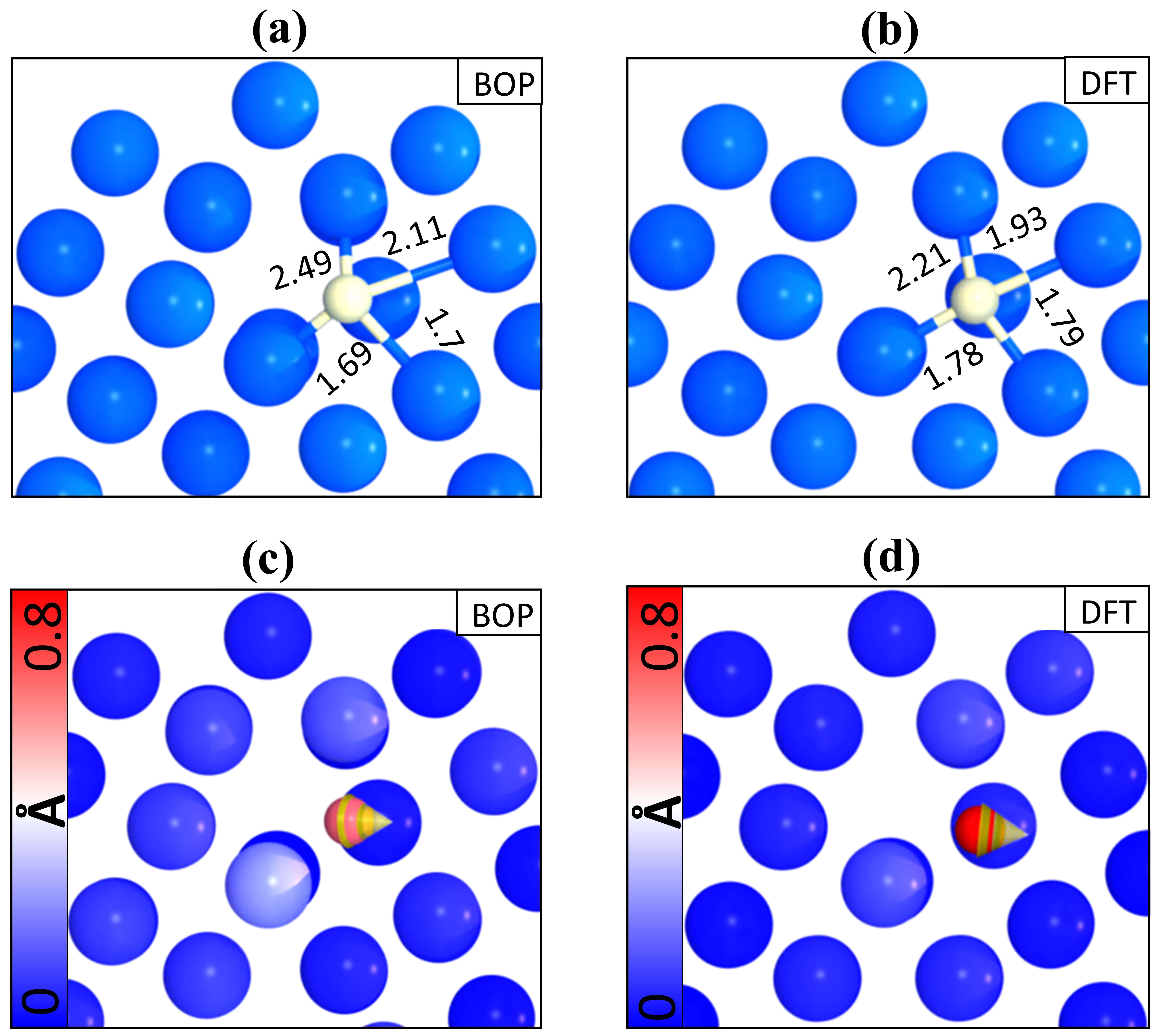}
\caption{(a) Relaxed configuration of the lowest energy interstitial H site at Cu GB obtained using BOP; (b) DFT. The distances in \AA~between H and neighboring Cu atoms are included. Cu and H atoms are shown in blue and white, respectively. (c) Displacements of atoms obtained using DFT; (d) BOP. Atoms are colored based on the displacement magnitude. Yellow arrows illustrate the displacement vectors. The initial configuration prior to relaxation is used as a reference.}
  \label{fig:relaxedconfigurations}
\end{figure}

\subsection{Comparison of DFT and BOP calculations}

One of the goals of this study is to compare the DFT findings to those obtained using BOP. The main energetic properties that we are interested in are the segregation and strengthening energies of H in Cu GB along with the induced lattice relaxation effects. Figure \ref{fig:hingrainboundaries} summarizes the segregation results for H in Cu GB obtained using DFT and BOP. The comparison of the segregation energies of H between the two methods for all examined interstitial sites is shown in Figure \ref{fig:hingrainboundaries}(b). H prefers to segregate as an interstitial atom instead of substitutional, with the octahedral interstitial position being the most favorable in the bulk. The octahedral interstitial site is more favorable compared to the tetrahedral one using DFT and BOP by 0.32 and 0.2\,eV, respectively. As seen in Figure \ref{fig:hingrainboundaries}(b), both DFT and BOP identify interstitial site 2 as the most favorable segregation site. The two methods give segregation energies within 0.26\,eV difference. All interstitial sites show negative segregation energies (favorable GB segregation). 

The two methods are also in good agreement regarding the strengthening energy (Figure \ref{fig:hingrainboundaries}(b)). In all four tested interstitial segregation sites, H is found to be more favorable to segregate in a hollow site of the (100) surface instead of any of the four examined interstitial sites in the $\Sigma$5 GB (positive strengthening energy). This indicates that the presence of H in the GBs will cause local decohesion, since it will facilitate the formation of a free surface in between the two grains. BOP and DFT give strengthening energies within 0.4\,eV difference, with BOP giving consistently lower energies. It can be deduced that, although there is a non-negligible discrepancy between the two methods, the computed energies follow a similar pattern at different segregation sites.

Figures \ref{fig:relaxedconfigurations}(a) and (b) illustrate the fully relaxed configuration of H introduced in the most favorable interstitial site (site 2 as seen in Figure \ref{fig:hingrainboundaries}(a)(i)) using BOP and DFT, respectively. The resulting distances between H and the four neighboring Cu atoms obtained with both methods agree within 5$\%$ of Cu's lattice constant, namely 3.62\,${\AA}$. Also, Figures \ref{fig:relaxedconfigurations}(b) and (c) illustrate the displacement vectors (yellow arrows) while atoms are colored based on their displacement, using the initial configuration prior to relaxation as a reference. Both DFT and BOP showed that the most significant displacement is that by H atom by approximately 0.8\,${\AA}$,  which during relaxation causes the displacements by around 0.1\,${\AA}$ of the four adjacent Cu atoms.

\begin{figure}[t]\vspace*{4pt}
 \centering
   \includegraphics[scale=0.034]{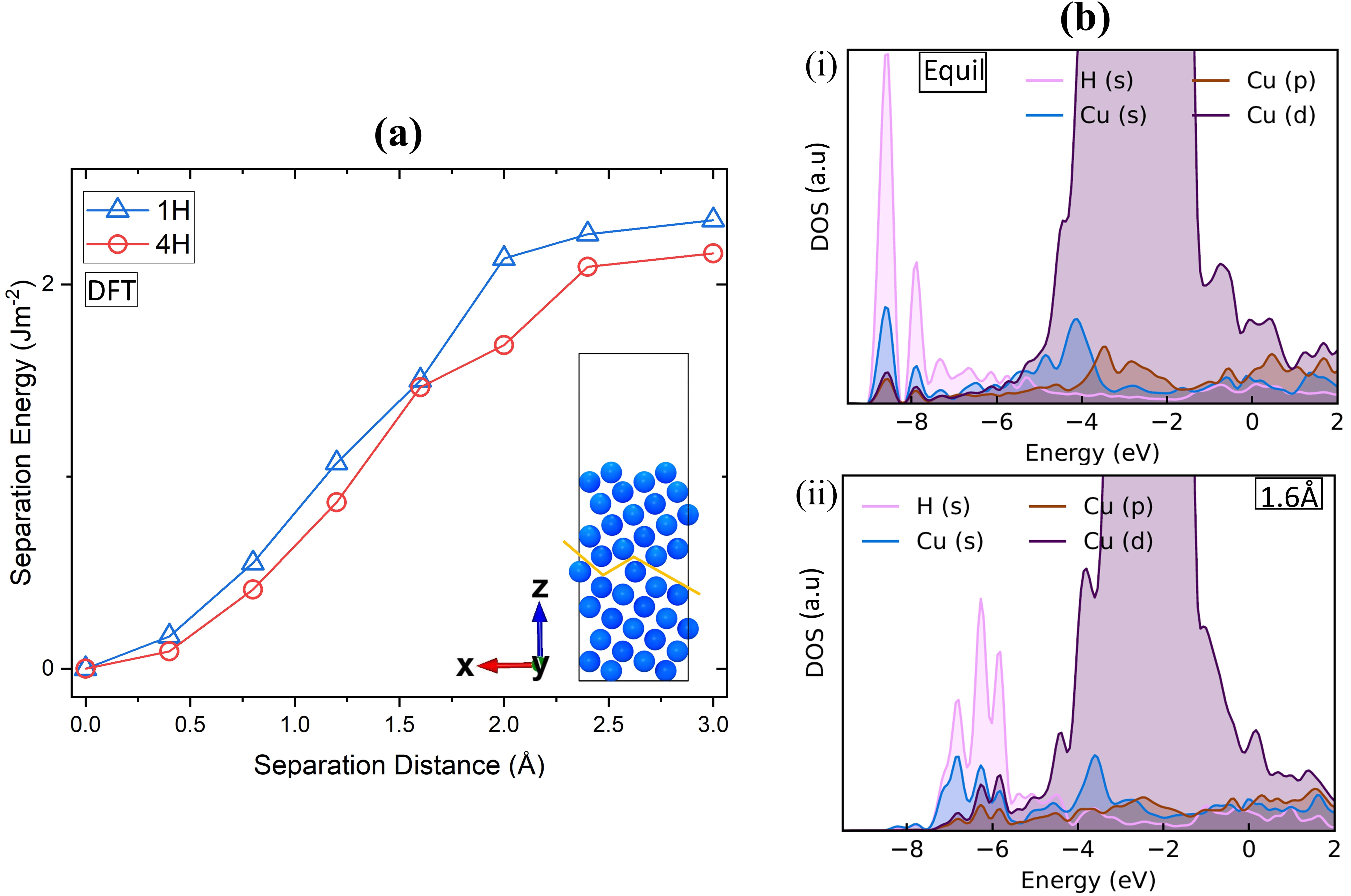}\\
\caption{(a) Separation energies of Cu GB with one and four interstitial H atoms at various separation distances from the equilibrium computed using DFT. Inset image illustrates the GB simulation cell along with the chosen separation plane (yellow line). (b) PDOS of the Cu1 (as seen in Figure \ref{fig:charges}(a)) and H atoms in the H-segregated GB model at 
(i) the equilibrium separation and (ii) a separation of 1.6\,\AA. The Fermi level is defined as the zero of energy. Hybridization peaks are observed at -9\,eV to -7\,eV between the Cu1 (see Figure \ref{fig:charges}(a)) and H atoms for the equilibrium separation and at -8\,eV to -5\,eV for the separation of 1.6\,\AA.}
  \label{fig:separation}
\end{figure}

\subsection{Ab initio strain calculations}

So far our results suggest that BOP and DFT are broadly in a good agreement. Prior to conducting MD simulations employing BOP, we sought to understand the underlying mechanism behind the observed weakening effect of H when introduced into Cu GB. Figure \ref{fig:separation}(a) includes the separation energy of the GB with 1 and 4 H atoms occupying the identified most favorable interstitial segregation sites (site 2 as seen in Figure \ref{fig:hingrainboundaries}(a)(i)). The inset image illustrates the GB simulation cell along with the plane chosen for the separation, shown as a yellow line. The plot includes full relaxation results obtained using DFT. The GB separation is increased up to the point where the grains are separated and a free surface is formed. The results follow the universal binding energy relation~\cite{ferrante1981universal}, increasing rapidly at the beginning and at larger distances start reaching an asymptote. At small distances, also referred to as pre-fracture region~\cite{huang2018uncovering}, the rearrangement of atoms heals the separation between the two grains. The difference between the energies of the separated grains and the GB at the equilibrium separation corresponds to the separation energy. It can be seen that, as the number of H atoms in the GB increases, the separation energy is reduced by approximately 0.1--0.3\,Jm$^{-2}$. A similar reduction in separation energy due to the presence of H has been reported in Ni GBs~\cite{mai2021understanding}. Also, we expect these effects to increase with the addition of more H atoms, as shown by~\cite{yamaguchi2019first} where up to 14 H interstitials in Al GBs led to a constant decrease in the cohesive energy. These results also are in good agreement with the computed strengthening energies shown in Figure \ref{fig:hingrainboundaries}(b), where the presence of H shows a weakening effect in the GB.

\begin{figure}[t]\vspace*{4pt}
 \centering
 
 \includegraphics[scale=0.025]{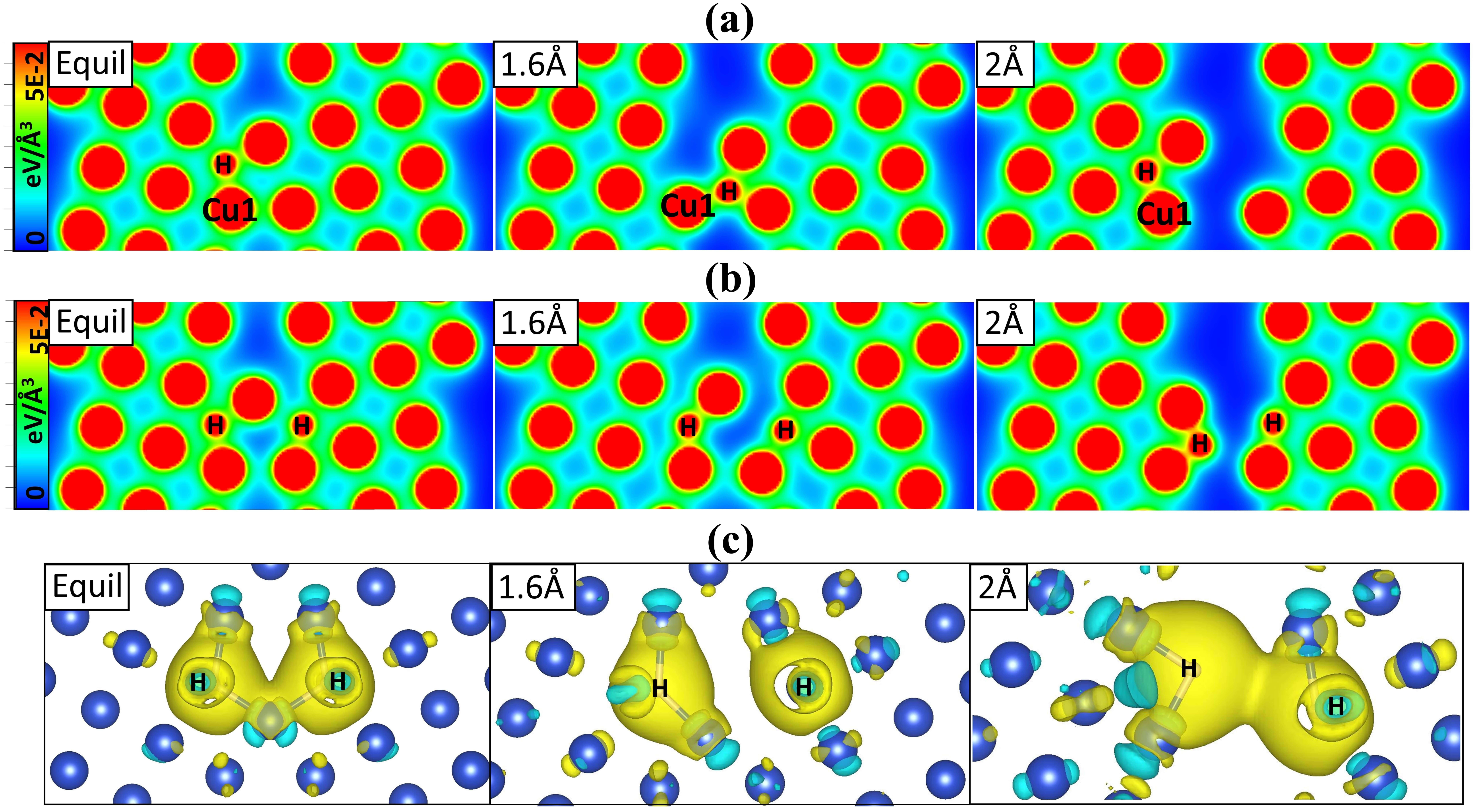}\\

\caption{Total charge distributions obtained using DFT at equilibrium (first column), 1.6\,{\AA} (second column), and 2\,{\AA} (third column) separations for Cu GB with (a) one interstitial H (b) four interstitial H atoms. (c) Differential charge distribution of four H interstitial in Cu GB at three different separations, namely equilibrium (first column), 1.6\,{\AA} (second column), and 2\,{\AA} (third column). The GB separation distances are the same as in (a) and (b). Yellow and cyan iso-surfaces (0.01) correspond to electron accumulation and depletion, respectively.}
  \label{fig:charges}
\end{figure}

Figures \ref{fig:separation}(b)(i) and (ii) show the projected density of states (PDOS) of the H atom and the nearest neighboring Cu atom (Cu1 as seen in Figure \ref{fig:charges}(a)) at the equilibrium separation and a separation of 1.6\,\AA~in the H-segregated GB model, respectively. At the equilibrium separation, hybridization peaks are observed at -9\,eV to -7\,eV between the Cu1 and H atoms. The hybridization is present in both separations. However, as the separation distance increases, the interaction between Cu1--H orbitals weakens. Figure \ref{fig:charges} illustrates the total charge density of the relaxed configurations at three different separations, namely the equilibrium, 1.6 and 2\,$\AA$ obtained using DFT with one (Figure \ref{fig:charges}(a)) and four (Figure \ref{fig:charges}(b)) H interstitials in Cu GB. As can be seen, H reduces the charge density between neighboring Cu atoms. The GB appears to be more prone to separation as the number of H atoms increases. Previous DFT investigations of one H interstitial in the GB of Fe found similar total charge density patterns~\cite{yuasa2012enhanced}. Our results are also in agreement with the mentioned above study in Fe, showing that a single H interstitial relaxes at the center of symmetry of the GB for a separation of 2\,$\AA$. The differential charge density (Figure \ref{fig:charges}(c)) shows that H  tends to accumulate electrons, bonding with adjacent Cu atoms as a result of the hybridization of the s H orbital and s/p/d orbitals from the host Cu atoms. 

\subsection{MD strain calculations}

\begin{figure}[t]\vspace*{4pt}
 \centering
   \begin{tabular}{cc}
      \textbf{(a)} & \hspace{6mm}\textbf{(b)}\\
        \includegraphics[scale=0.026]{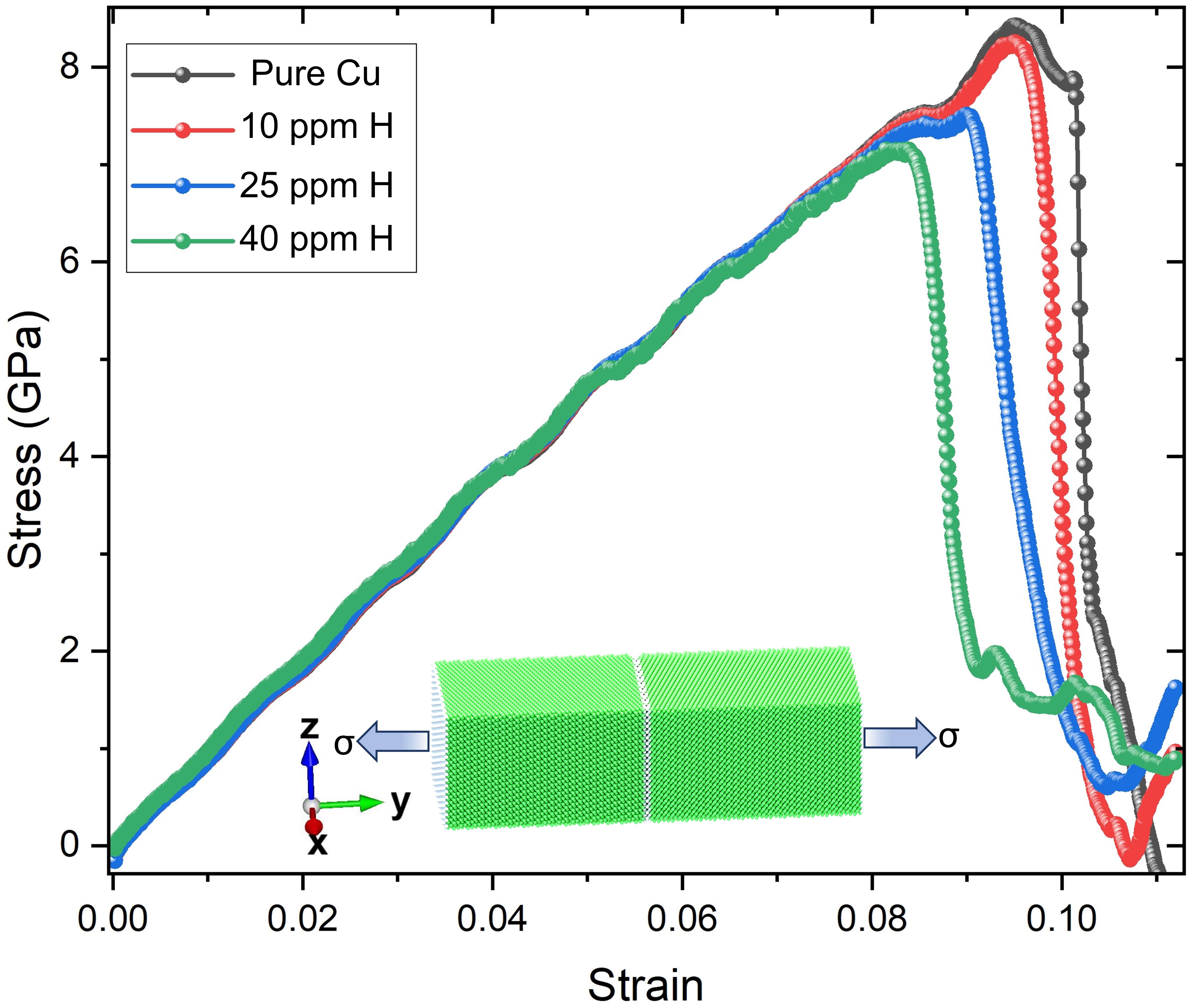}&
   \includegraphics[scale=0.28]{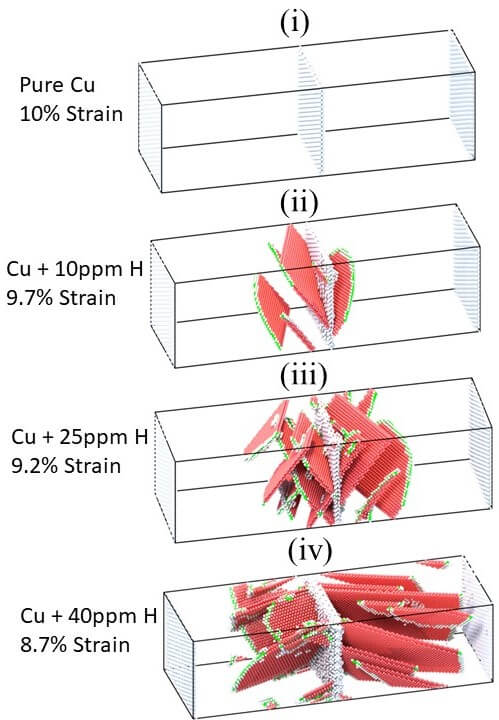}\\
 
\end{tabular}
\caption{(a) Stress-strain plots for pure Cu GB and Cu GB with different mass ppm concentrations of interstitial H atoms. Inset image illustrates the 120,000-atom Cu GB simulation cell. Green atoms correspond to fcc symmetry whereas amorphous regions (GB) are shown in white. Uniaxial deformation is applied along the y-axis. (b) Dislocation emission from the GB for (i) pure Cu and Cu with (ii) 10 (ii) 25 (iii) 40 mass ppm of H interstitials. Only atoms with non-fcc symmetry are visualized, with red and white corresponding to atoms with hcp symmetry and amorphous structure (GB), respectively. Green lines correspond to Shockley dislocations. Visualization is performed using OVITO~\cite{stukowski2009visualization}.}
  \label{fig:tensile}
\end{figure}

Our results predict that the presence of H in Cu GB can have a significant brittle effect. Hence, it is of interest to understand the mechanical effect of H when introduced in Cu GB along with the concentrations needed for such an effect to be significant. First, we introduce H interstitials at the  identified most favorable sites (site 2 as seen in Figure \ref{fig:hingrainboundaries}(a)(i)) and at various concentrations in the 120,000-atom bicrystalline $\Sigma5$ GB simulation cell shown in Figure \ref{fig:tensile}(a) (inset). Once the cells with and without H are fully relaxed and equilibrated, uniaxial tensile strain is applied along the y-axis.

Figure \ref{fig:tensile}(a) shows the effect of H on the mechanical strength of bicrystalline Cu at increasing mass ppm concentrations (10--40 mass ppm). As can be deduced, for any H concentration higher than 10 mass ppm in Cu GB, the yield strength of the crystal is significantly reduced, namely from 8.4\,GPa (pure Cu) to 7.3\,GPa (Cu with 40 mass ppm H). Figures \ref{fig:tensile}(b)(i)--(iv)  illustrate the correlation between formed partial Shockley dislocations and the concentration of H. The formation of hcp planes (red atoms in Figure \ref{fig:tensile}(b)) from the GB initiate as Shockley dislocations (green lines as seen in Figure \ref{fig:tensile}(b)). Therefore, it is apparent that the presence of H interstitials in Cu GB increases the density of partial Shockley dislocations. The latter leads to a significant decrease in the yield strength of the bicrystal. A similar increase in the density of dislocations has been observed in previous MD tensile strain calculations in $\alpha$-Fe~\cite{wan2019hydrogen}. However, we should note that the H solubility in Fe~\cite{nazarov2010first} is considerably higher compared to Cu~\cite{fukai2003superabundant,fukumuro2011influence}. Therefore, obtaining such H concentrations in ECD Cu is unlikely.

\section{Conclusions}

We conducted DFT and BOP MD simulations to understand the mechanical effect of the inclusion of H interstitials in Cu GB. Using DFT, H was found to reduce the separation energy of Cu $\Sigma$5 GB. This effect was attributed to the induced lattice relaxation and local charge redistribution caused by H, which reduced the charge density between neighboring Cu atoms. The latter led to a weakening of the Cu--Cu bonds close to the H interstitials, causing a local weakening effect.

Compared to the DFT results, BOP accurately predicts the most favorable H segregation sites in Cu GB. Also, BOP calculations are in broadly good agreement with DFT regarding the strengthening energy and local relaxation. Using BOP, H was found to facilitate the emission of Shockley and partial dislocations at the GB. However, at least 10 mass ppm of H was needed for a substantial drop in the yield strength to occur. Further study will be needed to understand the effect of H on the mechanical properties of Cu while taking into account the presence of triple junctions.

\section*{Acknowledgements}

A.L.S. acknowledges funding by EPSRC (grant EP/P013503/1). V.F. would like to acknowledge funding by EPSRC (grant EP/L015862/1) as part of the CDT in molecular modeling and materials science. Computational resources on ARCHER2 (http://www.archer2.ac.uk) were provided via our membership in the UK's HPC Materials Chemistry Consortium, which is funded by EPSRC (EP/L000202, EP/R029431). V.F. and A.L.S. would like to thank Jack Strand, Thomas Durrant, Ricardo Grau-Crespo and Corey S. O'Hern for useful comments and help in calculations.





\bibliographystyle{unsrt}
\bibliography{xampl}

\begin{thebibliography}{10}

\bibitem{troiano1960role}
Alexander~R Troiano.
\newblock The role of hydrogen and other interstitials in the mechanical
  behavior of metals.
\newblock {\em trans. ASM}, 52:54--80, 1960.

\bibitem{louthan2008hydrogen}
MR~Louthan.
\newblock Hydrogen embrittlement of metals: a primer for the failure analyst.
\newblock {\em Journal of Failure Analysis and Prevention}, 8(3):289--307,
  2008.

\bibitem{djukic2016hydrogen}
Milos~B Djukic, Gordana~M Bakic, Vera~Sijacki Zeravcic, Aleksandar Sedmak, and
  Bratislav Rajicic.
\newblock Hydrogen embrittlement of industrial components: prediction,
  prevention, and models.
\newblock {\em Corrosion}, 72(7):943--961, 2016.

\bibitem{li2019effects}
Jiaqing Li, Cheng Lu, Linqing Pei, Che Zhang, Rui Wang, and Kiet Tieu.
\newblock Effects of {H} segregation on shear-coupled motion of 〈110〉 grain
  boundaries in $\alpha$-{F}e.
\newblock {\em international journal of hydrogen energy}, 44(18616):e18627,
  2019.

\bibitem{moser2021electropolishing}
Sebastian Moser, Manuel Kleinbichler, Sabine Kubicek, Johannes Zechner, and
  Megan~J Cordill.
\newblock Electropolishing--a practical method for accessing voids in metal
  films for analyses.
\newblock {\em Applied Sciences}, 11(15):7009, 2021.

\bibitem{konishi2002effect}
Shinya Konishi, Miki Moriyama, and Masanori Murakami.
\newblock Effect of annealing atmosphere on void formation in copper
  interconnects.
\newblock {\em Materials Transactions}, 43(7):1624--1628, 2002.

\bibitem{yamaguchi2019first}
M~Yamaguchi, K-I Ebihara, M~Itakura, T~Tsuru, K~Matsuda, and HJCMS Toda.
\newblock First-principles calculation of multiple hydrogen segregation along
  aluminum grain boundaries.
\newblock {\em Computational Materials Science}, 156:368--375, 2019.

\bibitem{lousada2022hydrogen}
Cl{\'a}udio~M Lousada and Pavel~A Korzhavyi.
\newblock Hydrogen at symmetric tilt grain boundaries in aluminum: segregation
  energies and structural features.
\newblock {\em Scientific Reports}, 12(1):19872, 2022.

\bibitem{fotopoulos2023molecular}
Vasileios Fotopoulos, Corey~S O'Hern, and Alexander~L Shluger.
\newblock Molecular dynamics simulations of the thermal evolution of voids in
  {C}u bulk and grain boundaries.
\newblock In {\em TMS 2023 152nd Annual Meeting \& Exhibition Supplemental
  Proceedings}, pages 1001--1010. Springer, 2023.

\bibitem{wan2019hydrogen}
Liang Wan, Wen~Tong Geng, Akio Ishii, Jun-Ping Du, Qingsong Mei, Nobuyuki
  Ishikawa, Hajime Kimizuka, and Shigenobu Ogata.
\newblock Hydrogen embrittlement controlled by reaction of dislocation with
  grain boundary in alpha-iron.
\newblock {\em International Journal of Plasticity}, 112:206--219, 2019.

\bibitem{lynch2012hydrogen}
Stan Lynch.
\newblock Hydrogen embrittlement phenomena and mechanisms.
\newblock {\em Corrosion Reviews}, 30(3-4):105--123, 2012.

\bibitem{birnbaum1994hydrogen}
Howard~K Birnbaum and Petros Sofronis.
\newblock Hydrogen-enhanced localized plasticity--a mechanism for
  hydrogen-related fracture.
\newblock {\em Materials Science and Engineering: A}, 176(1-2):191--202, 1994.

\bibitem{robertson2001effect}
IM~Robertson.
\newblock The effect of hydrogen on dislocation dynamics.
\newblock {\em Engineering fracture mechanics}, 68(6):671--692, 2001.

\bibitem{articleNeedleman}
Alan Needleman.
\newblock A continuum model for void nucleation by inclusion debonding.
\newblock {\em Journal of Applied Mechanics}, 54, 09 1987.

\bibitem{gabe1997role}
DR~Gabe.
\newblock The role of hydrogen in metal electrodeposition processes.
\newblock {\em Journal of applied electrochemistry}, 27(8):908--915, 1997.

\bibitem{li2020review}
Xinfeng Li, Xianfeng Ma, Jin Zhang, Eiji Akiyama, Yanfei Wang, and Xiaolong
  Song.
\newblock Review of hydrogen embrittlement in metals: Hydrogen diffusion,
  hydrogen characterization, hydrogen embrittlement mechanism and prevention.
\newblock {\em Acta Metallurgica Sinica (English Letters)}, 33:759--773, 2020.

\bibitem{wipf1997diffusion}
H~Wipf.
\newblock Diffusion of hydrogen in metals.
\newblock {\em Hydrogen in metals III}, pages 51--91, 1997.

\bibitem{fotopoulos2023thermodynamic}
Vasileios Fotopoulos, Ricardo Grau-Crespo, and Alexander Shluger.
\newblock Thermodynamic analysis of the interaction between metal vacancies and
  hydrogen in bulk {C}u.
\newblock {\em Physical Chemistry Chemical Physics}, 25:9168--9175, 2023.

\bibitem{yoo2021density}
SangHyuk Yoo, Byeongchan Lee, and Keonwook Kang.
\newblock Density functional theory study of the mechanical behavior of
  silicene and development of a {T}ersoff interatomic potential model tailored
  for elastic behavior.
\newblock {\em Nanotechnology}, 32(29):295702, 2021.

\bibitem{wang2013defect}
Hao Wang, David Rodney, DS~Xu, Rui Yang, and P~Veyssi{\`e}re.
\newblock Defect kinetics on experimental timescales using atomistic
  simulations.
\newblock {\em Philosophical Magazine}, 93(1-3):186--202, 2013.

\bibitem{zhou2015analytical}
XW~Zhou, DK~Ward, M~Foster, and JA~Zimmerman.
\newblock An analytical bond-order potential for the copper--hydrogen binary
  system.
\newblock {\em Journal of Materials Science}, 50(7):2859--2875, 2015.

\bibitem{zhou2018bond}
XW~Zhou, DK~Ward, and ME~Foster.
\newblock A bond-order potential for the {A}l--{C}u--{H} ternary system.
\newblock {\em New Journal of Chemistry}, 42(7):5215--5228, 2018.

\bibitem{drautz2015bond}
Ralf Drautz, Thomas Hammerschmidt, Miroslav {\v{C}}{\'a}k, and DG~Pettifor.
\newblock Bond-order potentials: derivation and parameterization for refractory
  elements.
\newblock {\em Modelling and Simulation in Materials Science and Engineering},
  23(7):074004, 2015.

\bibitem{silver1996kernel}
RN~Silver, H~Roeder, AF~Voter, and JD~Kress.
\newblock Kernel polynomial approximations for densities of states and spectral
  functions.
\newblock {\em Journal of Computational Physics}, 124(1):115--130, 1996.

\bibitem{voter1996linear}
AF~Voter, JD~Kress, and RN~Silver.
\newblock Linear-scaling tight binding from a truncated-moment approach.
\newblock {\em Physical Review B}, 53(19):12733, 1996.

\bibitem{oleinik1999analytic}
II~Oleinik and DG~Pettifor.
\newblock Analytic bond-order potentials beyond {T}ersoff-{B}renner. {II}.
  {A}pplication to the hydrocarbons.
\newblock {\em Physical Review B}, 59(13):8500, 1999.

\bibitem{juslin2005analytical}
Niklas Juslin, P~Erhart, Petra Tr{\"a}skelin, Janne Nord, Krister~OE
  Henriksson, Kai Nordlund, Emppu Salonen, and K~Albe.
\newblock Analytical interatomic potential for modeling nonequilibrium
  processes in the {W}--{C}--{H} system.
\newblock {\em Journal of applied physics}, 98(12):123520, 2005.

\bibitem{kresse1993ab}
Georg Kresse and JJPRB Hafner.
\newblock Ab initio molecular dynamics for open-shell transition metals.
\newblock {\em Physical Review B}, 48(17):13115, 1993.

\bibitem{kresse1996efficient}
Georg Kresse and J{\"u}rgen Furthm{\"u}ller.
\newblock Efficient iterative schemes for ab initio total-energy calculations
  using a plane-wave basis set.
\newblock {\em Physical review B}, 54(16):11169, 1996.

\bibitem{kresse1996efficiency}
Georg Kresse and J{\"u}rgen Furthm{\"u}ller.
\newblock Efficiency of ab-initio total energy calculations for metals and
  semiconductors using a plane-wave basis set.
\newblock {\em Computational materials science}, 6(1):15--50, 1996.

\bibitem{perdew1996generalized}
John~P Perdew, Kieron Burke, and Matthias Ernzerhof.
\newblock Generalized gradient approximation made simple.
\newblock {\em Physical Review Letters}, 77(18):3865, 1996.

\bibitem{ganchenkova2014effects}
MG~Ganchenkova, YN~Yagodzinskyy, VA~Borodin, and Hannu H{\"a}nninen.
\newblock Effects of hydrogen and impurities on void nucleation in copper:
  simulation point of view.
\newblock {\em Philosophical Magazine}, 94(31):3522--3548, 2014.

\bibitem{er1975iterativecalculationof}
ER~Davidson.
\newblock The iterative calculation of a few of the lowest eigenvalues and
  corresponding eigenvectors of large real-symmetric matrices.
\newblock {\em Journal of Computational Physics}, 17:87--94, 1975.

\bibitem{pulay1980convergence}
P{\'e}ter Pulay.
\newblock Convergence acceleration of iterative sequences. the case of {SCF}
  iteration.
\newblock {\em Chemical Physics Letters}, 73(2):393--398, 1980.

\bibitem{wood1985new}
DM~Wood and Alex Zunger.
\newblock A new method for diagonalising large matrices.
\newblock {\em Journal of Physics A: Mathematical and General}, 18(9):1343,
  1985.

\bibitem{wu2016first}
Xuebang Wu, Yu-Wei You, Xiang-Shan Kong, Jun-Ling Chen, G-N Luo, Guang-Hong Lu,
  CS~Liu, and Zhiguang Wang.
\newblock First-principles determination of grain boundary strengthening in
  tungsten: Dependence on grain boundary structure and metallic radius of
  solute.
\newblock {\em Acta Materialia}, 120:315--326, 2016.

\bibitem{nazarov2012vacancy}
Roman Nazarov, Tilmann Hickel, and J{\"o}rg Neugebauer.
\newblock Vacancy formation energies in fcc metals: influence of
  exchange-correlation functionals and correction schemes.
\newblock {\em Physical Review B}, 85(14):144118, 2012.

\bibitem{nazarov2014ab}
Roman Nazarov, Tilmann Hickel, and J{\"o}rg Neugebauer.
\newblock Ab initio study of {H}-vacancy interactions in fcc metals:
  Implications for the formation of superabundant vacancies.
\newblock {\em Physical Review B}, 89(14):144108, 2014.

\bibitem{bodlos2022energies}
R~Bodlos, V~Fotopoulos, J~Spitaler, AL~Shluger, and L~Romaner.
\newblock Energies and structures of {C}u/{N}b and {C}u/{W} interfaces from
  density functional theory and semi-empirical calculations.
\newblock {\em Materialia}, 21:101362, 2022.

\bibitem{thompson2022lammps}
Aidan~P Thompson, H~Metin Aktulga, Richard Berger, Dan~S Bolintineanu,
  W~Michael Brown, Paul~S Crozier, Pieter~J in't Veld, Axel Kohlmeyer, Stan~G
  Moore, Trung~Dac Nguyen, et~al.
\newblock {LAMMPS}-a flexible simulation tool for particle-based materials
  modeling at the atomic, meso, and continuum scales.
\newblock {\em Computer Physics Communications}, 271:108171, 2022.

\bibitem{fukai2003superabundant}
Y~Fukai.
\newblock Superabundant vacancies formed in metal--hydrogen alloys.
\newblock {\em Physica Scripta}, 2003(T103):11, 2003.

\bibitem{fukumuro2011influence}
N~Fukumuro, T~Adachi, S~Yae, H~Matsuda, and Y~Fukai.
\newblock Influence of hydrogen on room temperature recrystallisation of
  electrodeposited {C}u films: thermal desorption spectroscopy.
\newblock {\em Transactions of the IMF}, 89(4):198--201, 2011.

\bibitem{zhao2016ductile}
Kai Zhao, Inga~Gudem Ringdalen, Jianyang Wu, Jianying He, and Zhiliang Zhang.
\newblock Ductile mechanisms of metals containing pre-existing nanovoids.
\newblock {\em Computational Materials Science}, 125:36--50, 2016.

\bibitem{zhou2017molecular}
Kai Zhou, Bin Liu, Shaofeng Shao, and Yijun Yao.
\newblock Molecular dynamics simulations of tension--compression asymmetry in
  nanocrystalline copper.
\newblock {\em Physics Letters A}, 381(13):1163--1168, 2017.

\bibitem{kapoor1998determination}
Rajeev Kapoor and Sia Nemat-Nasser.
\newblock Determination of temperature rise during high strain rate
  deformation.
\newblock {\em Mechanics of materials}, 27(1):1--12, 1998.

\bibitem{stukowski2012automated}
Alexander Stukowski, Vasily~V Bulatov, and Athanasios Arsenlis.
\newblock Automated identification and indexing of dislocations in crystal
  interfaces.
\newblock {\em Modelling and Simulation in Materials Science and Engineering},
  20(8):085007, 2012.

\bibitem{momma2008vesta}
Koichi Momma and Fujio Izumi.
\newblock Vesta: a three-dimensional visualization system for electronic and
  structural analysis.
\newblock {\em Journal of Applied Crystallography}, 41(3):653--658, 2008.

\bibitem{razumovskiy2018solute}
VI~Razumovskiy, SV~Divinski, and L~Romaner.
\newblock Solute segregation in {C}u: {DFT} vs. experiment.
\newblock {\em Acta Materialia}, 147:122--132, 2018.

\bibitem{wurmshuber2022mechanical}
Michael Wurmshuber, Michael Burtscher, Simon Doppermann, Rishi Bodlos, Daniel
  Scheiber, Lorenz Romaner, and Daniel Kiener.
\newblock Mechanical performance of doped {W}--{C}u nanocomposites.
\newblock {\em Materials Science and Engineering: A}, 857:144102, 2022.

\bibitem{rice1989embrittlement}
James~R Rice and Jian-Sheng Wang.
\newblock Embrittlement of interfaces by solute segregation.
\newblock {\em Materials Science and Engineering: A}, 107:23--40, 1989.

\bibitem{yamaguchi2005grain}
Masatake Yamaguchi, Motoyuki Shiga, and Hideo Kaburaki.
\newblock Grain boundary decohesion by impurity segregation in a nickel-sulfur
  system.
\newblock {\em Science}, 307(5708):393--397, 2005.

\bibitem{scheiber2020influence}
D~Scheiber, K~Prabitz, L~Romaner, and W~Ecker.
\newblock The influence of alloying on {Z}n liquid metal embrittlement in
  steels.
\newblock {\em Acta Materialia}, 195:750--760, 2020.

\bibitem{ferrante1981universal}
J~Ferrante, JR~Smith, and JH~Rose.
\newblock Universal binding energy relations in metallic adhesion.
\newblock In {\em Tribology Series}, volume~7, pages 19--30. Elsevier, 1981.

\bibitem{huang2018uncovering}
Zhifeng Huang, Fei Chen, Qiang Shen, Lianmeng Zhang, and Timothy~J Rupert.
\newblock Uncovering the influence of common nonmetallic impurities on the
  stability and strength of a (310) grain boundary in {C}u.
\newblock {\em Acta Materialia}, 148:110--122, 2018.

\bibitem{mai2021understanding}
Han~Lin Mai, Xiang-Yuan Cui, Daniel Scheiber, Lorenz Romaner, and Simon~P
  Ringer.
\newblock An understanding of hydrogen embrittlement in nickel grain boundaries
  from first principles.
\newblock {\em Materials \& Design}, 212:110283, 2021.

\bibitem{yuasa2012enhanced}
Motohiro Yuasa, Takashi Amemiya, and Mamoru Mabuchi.
\newblock Enhanced grain boundary embrittlement of an {F}e grain boundary
  segregated by hydrogen {(H)}.
\newblock {\em Journal of Materials Research}, 27(12):1589--1597, 2012.

\bibitem{nazarov2010first}
R~Nazarov, T~Hickel, and J~Neugebauer.
\newblock First-principles study of the thermodynamics of hydrogen-vacancy
  interaction in fcc iron.
\newblock {\em Physical Review B}, 82(22):224104, 2010.

\bibitem{stukowski2009visualization}
Alexander Stukowski.
\newblock Visualization and analysis of atomistic simulation data with
  {OVITO}--the open visualization tool.
\newblock {\em Modelling and Simulation in Materials Science and Engineering},
  18(1):015012, 2009.

\end{thebibliography}










\clearpage

\normalMode

\end{document}